\documentclass{article}
\usepackage{spconf,amsmath,graphicx}
\usepackage{hyperref}
\usepackage{multirow}
\usepackage{booktabs}
\usepackage{color}

\title{Spoofing-Aware Speaker Verification Robust Against Domain and Channel Mismatches}
\name{Chang Zeng$^{1,2}$, Xiaoxiao Miao$^3$, Xin Wang$^1$, Erica Cooper$^1$, Junichi Yamagishi$^{1,2}$}
\address{$^1$National Insitute of Informatics, Japan \; $^2$SOKENDAI, Japan \\ $^3$Singapore Institute of Technology, Singapore}
\begin{document}
\ninept
\maketitle
\begin{abstract}
In real-world applications, it is challenging to build a speaker verification system that is simultaneously robust against common threats, including spoofing attacks, channel mismatch, and domain mismatch. Traditional automatic speaker verification (ASV) systems often tackle these issues separately, leading to suboptimal performance when faced with simultaneous challenges. In this paper, we propose an integrated framework that incorporates pair-wise learning and spoofing attack simulation into the meta-learning paradigm to enhance robustness against these multifaceted threats. This novel approach employs an asymmetric dual-path model and a multi-task learning strategy to handle ASV, anti-spoofing, and spoofing-aware ASV tasks concurrently. A new testing dataset, CNComplex, is introduced to evaluate system performance under these combined threats. Experimental results demonstrate that our integrated model significantly improves performance over traditional ASV systems across various scenarios, showcasing its potential for real-world deployment. Additionally, the proposed framework's ability to generalize across different conditions highlights its robustness and reliability, making it a promising solution for practical ASV applications.
\end{abstract}
\begin{keywords}
Speaker verification, robustness, multi-task learning, meta-learning
\end{keywords}
\section{Introduction}
\label{sec:introduction}
Automatic speaker verification (ASV) aims to accurately identify the speaker's identity by analyzing voice characteristics within speech signals, providing robust support for personalized services. Mainstream ASV solutions have evolved from traditional statistics-based methods such as GMM-UBM \cite{gmmubm} and I-Vector \cite{ivector} to deep neural network-based approaches \cite{xvector1, resnet34-fast}. For example, time-delay neural network (TDNN) \cite{tdnn,ftdnn,ecapa-tdnn} and ResNet \cite{resnet,resnet34-performance} have dominated the realm of extracting speaker embeddings since they can efficiently leverage large amounts of speech data to capture the properties of speakers accurately. Additionally, the back-end is another important component of an ASV system. Neural network-based methods \cite{nplda,attention-backend,zeng2024joint} have gradually replaced the widely used PLDA model in recent years.

\begin{figure}[tbp]
\centering
\includegraphics[width=8cm]{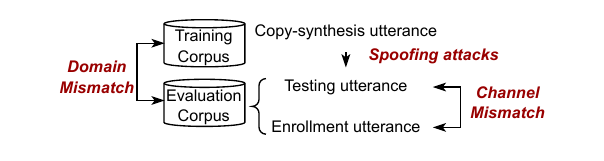}
\caption{The primary threats to ASV systems. These threats can be categorized into channel mismatch, spoofing attacks, and domain mismatch.}
\label{fig:threats}
\vspace{-4mm}
\end{figure}

However, as speaker recognition has rapidly evolved, its focus has shifted from improving performance on simple datasets such as TIMIT and LibriSpeech \cite{timit,librispeech} to addressing challenges in what is often referred to as ``recognition in the wild'' \cite{voxceleb1,sitw,cnceleb1}. In this context, numerous issues have surfaced, including channel mismatch \cite{gplda1,plda,gplda2,normdetect}, vulnerability to spoofing attacks \cite{asvspoof2019,asvspoof2021,add2022}, and domain mismatches \cite{domain-invariant}, as shown in Figure \ref{fig:threats}. While substantial progress has been made in addressing each of these challenges individually \cite{channelmismatch1,sasv-attention-backend,robustmaml}, the practical deployment of ASV systems demands solutions that can effectively handle these multifaceted threats. Given the potential for the ASV system to encounter simultaneous attacks, there is a critical need to amalgamate all these functionalities within the ASV system itself.

However, the integration is hindered by two factors.
\begin{enumerate}
    \item Presently, no dataset encompasses the presence of three concurrent threats. For instance, widely used datasets for speaker verification such as VoxCeleb 1\&2 \cite{voxceleb1,voxceleb2} and CNCeleb 1\&2 \cite{cnceleb1,cnceleb2} exclusively feature scenarios involving channel mismatch. Meanwhile, datasets designed for the ASVspoof challenge series \cite{asvspoof2019,asvspoof2021} solely tackle spoofing attacks. Hence, it is imperative to develop a new dataset and evaluation protocol encompassing all three considerations.
    \item One apparent method to incorporate all functionalities involves cascading individual components, each dedicated to addressing a specific challenge. For instance, positioning the countermeasures module ahead of the ASV module serves to detect synthesized input utterances. Additionally, regarding the ASV module as comprising a neural speaker encoder and a back-end, assigning domain mismatch handling to the former and channel mismatch to the latter seems intuitive. However, as discussed in \cite{sasv-attention-backend}, this cascading approach overlooks the interactions among these modules, leading to suboptimal system performance. Consequently, a novel framework must be devised to facilitate and leverage these interactions effectively.
\end{enumerate}

In response to this practical need, in this paper we introduce an integrated framework that harmonizes the techniques elucidated in previous studies \cite{attention-backend,sasv-attention-backend,robustmaml}. This includes the pair-wise learning paradigm, which simulates channel mismatch, the fusion module, which replicates spoofing attacks, and the meta-learning paradigm, which mimics domain mismatch. The aim is to establish a robust defense against channel variability, spoofing attacks, and domain mismatch, all in one comprehensive approach. This represents an end-to-end methodology, optimizing resilience across these challenges by utilizing a shared model. To the best of our knowledge, this is the first attempt to integrate robustness into a single model by explicitly exposing the model to all threats. The primary contribution lies in amalgamating the pair-wise learning, spoofing simulation, and meta-learning paradigms into a unified optimization strategy.

The subsequent sections of this paper are structured as follows. We commence with a preliminary experiment with the ECAPA-TDNN model on a new spoofing-aware ASV evaluation dataset, which contains testing trials for both channel and domain mismatches, offering evidence of the vulnerability of contemporary SOTA ASV models when facing three threats simultaneously. Next, an overview of the proposed integrated approach is given in Section \ref{sec:integration}, including the overview of it to offer insights into the integration methodology in Section \ref{sec:overview}, the specifics of the model architecture and its objectives in Section \ref{sec:model}, and the learning paradigm for E2E optimization in Section \ref{sec:paradigm}. The experimental setup, along with the results and their analysis, is detailed in Section \ref{sec:exp}. Section \ref{sec:conclusion} concludes the paper with a comprehensive summary.

\section{Vulnerability of contemporary ASV model to simultaneous threats}
\label{sec:preliminary}

\begin{table}[tbp]
  \caption{Genre group division. Ten genres are randomly divided into four groups.}
  \vspace{2mm}
  \label{tab:genre-group}
  \centering
  \begin{tabular}{lc}
    \toprule
    \multicolumn{1}{l}{\textbf{Group}} & \multicolumn{1}{c}{\textbf{Genre Types}} \\
    \midrule
    Group I & drama (dr), vlog (vl), speech (sp) \\
    \midrule
    Group II & entertainment (en), interview (in), play (pl) \\
    \midrule
    Group III & live broadcast (lb), movie (mo) \\
    \midrule
    Group IV & singing (si), recitation (re) \\
    \bottomrule
  \end{tabular}
\end{table}

\begin{table}[tbp]
\setlength\tabcolsep{4pt}
  \caption{Cross-genre protocols (CGPs) established via K-fold validation. For each protocol, the training dataset only contains seen genre groups, while the testing dataset contains both seen and unseen genre groups. The unseen genre in the testing dataset results in a cross-genre scenario. Note that speakers in the training and testing datasets overlap.}
  \vspace{2mm}
  \label{tab:protocol}
  \centering
  \begin{tabular}{lccc}
    \toprule
    \multicolumn{1}{l}{\textbf{CGP}} & \multicolumn{1}{c}{\textbf{Seen Genres}} & \multicolumn{1}{c}{\textbf{Unseen Genres}} \\
    \midrule
    CGP I & Group I, Group II, Group III & Group IV \\
    \midrule
    CGP II & Group I, Group II, Group IV & Group III \\
    \midrule 
    CGP III & Group I, Group III, Group IV & Group II \\
    \midrule
    CGP IV & Group II, Group III, Group IV & Group I \\
    \bottomrule
  \end{tabular}
\end{table}

\subsection{Testing dataset}

In light of the absence of an existing dataset that aligns with requirements for evaluating system performance in the challenging scenario involving all three threats simultaneously, it is imperative to create a new testing dataset and devise a corresponding evaluation protocol tailored to this scenario. This newly developed test dataset was constructed based on the original CNCeleb test dataset, as CNCeleb has multiple genres, making it well-suited as a basis for building complex test sets. Within this dataset, the enrollment utterances remain unchanged. However, the testing utterances are subject to random substitution with re-vocoded data sourced from the CNSpoof dataset \cite{zc-meta-grl}. The new testing dataset is referred to as CNComplex.

In terms of establishing the ground truth, we adhere to the definition outlined in SASVC 2022 \cite{jung22sasvc}. Specifically, if the testing utterance is genuine and its corresponding speaker label matches the enrolled identity, it is labeled as true; otherwise, it is marked as false. Consequently, the new testing dataset now encompasses both channel mismatch and spoofing attacks.

\subsection{Preliminary experimental result of contemporary ASV model}
\label{sec:preliminary_result}

To demonstrate the vulnerability of the contemporary ASV model when faced with the CNComplex testing dataset, an ECAPA-TDNN speaker encoder was trained, incorporating the AM-Softmax loss function \cite{lmcl,amsoftmax-face}, on two separate datasets: CNCeleb 1\&2 and the combination of CNCeleb 1\&2 and CNSpoof datasets according to the cross-genre protocols (CGPs) described in \cite{zc-meta-grl}. Each protocol excludes data with certain genres from the training dataset. In detail, the ten genres present in the training data were categorized into four groups, as outlined in Table \ref{tab:genre-group}. It is worth noting that the "advertisement" genre was excluded from the dataset due to its limited amount of data. For each CGP, $660,000$ utterances were randomly sampled to compile a training dataset, which encompassed three of the genre groups. Importantly, in this way, each training dataset has an unseen genre group in the testing dataset. This strategy is detailed in Table \ref{tab:protocol}.

The results of this experiment evaluated by cosine similarity are presented in Table \ref{tab:result_ecapa}. We subjected these well-trained models to evaluation on both the original CNCeleb testing dataset (CNCeleb.Eval) and the newly introduced CNComplex dataset.

\begin{table}[tbp]
\footnotesize
\setlength\tabcolsep{0.8mm}
  \caption{Performance of the ECAPA-TDNN model trained on two distinct sets of data: CNCeleb 1\&2 and the combination of (CNCeleb 1\&2, CNSpoof) according to CGP. Subsequently, evaluations on both the original CNCeleb testing dataset and the newly developed testing dataset were conducted. The evaluation metrics employed for this assessment are the SV-EER and SASV-EER defined in \cite{sasv-attention-backend}.}
  \vspace{2mm}
  \label{tab:result_ecapa}
  \centering
  \begin{tabular}{lccccc}
    \toprule
    \multirow{2}{*}{\textbf{Protocol}} & \multirow{2}{*}{\textbf{Training dataset}} & \multicolumn{2}{c}{\textbf{CNCeleb.Eval}} & \multicolumn{2}{c}{\textbf{CNComplex}} \\
    \cmidrule(l{0em}r{0em}){3-6}
    & & SV-EER & SASV-EER & SV-EER & SASV-EER \\
    \midrule
    \multirow{2}{*}{\textbf{CGP I}}   & CNCeleb 1\&2             & 9.38  & - & 10.05 & 37.64  \\
                                      & Combination   & 9.02  & - & 9.56  & 36.97  \\
    \toprule
    \multirow{2}{*}{\textbf{CGP II}}  & CNCeleb 1\&2             & 10.04 & - & 10.85 & 40.76  \\
                                      & Combination   & 9.75  & - & 10.79 & 40.17  \\
    \toprule
    \multirow{2}{*}{\textbf{CGP III}} & CNCeleb 1\&2             & 10.12 & - & 10.67 & 39.62  \\
                                      & Combination   & 9.33  & - & 10.22 & 38.49  \\
    \toprule
    \multirow{2}{*}{\textbf{CGP IV}}  & CNCeleb 1\&2             & 9.59  & - & 10.24 & 37.97  \\
                                      & Combination   & 9.21  & - & 9.86  & 37.42  \\
    \bottomrule
  \end{tabular}
\end{table}

First and foremost, it is noteworthy that the ECAPA-TDNN model trained on the combination of CNCeleb 1\&2 and CNSpoof datasets exhibited a slightly consistent performance improvement on the CNCeleb.Eval dataset compared to that of the model trained solely on the CNCeleb 1\&2 dataset under all CGPs. This observation suggests that the copy-synthesis method \cite{copy-synthesis,copy-synthesis2} employed to generate the CNSpoof dataset can be considered as an effective data augmentation strategy in training an ASV model.

However, the critical focus shifts to the results obtained from the CNComplex dataset. As anticipated, the performance substantially deteriorated on this new dataset under all CGPs, regardless of which training dataset was used. This unequivocally indicates that the contemporary state-of-the-art ASV model is ill-equipped to address the complexities inherent in the new testing dataset, which contains simultaneous threats.

\section{Proposed integrated approach}
\label{sec:integration}

\subsection{Overview}
\label{sec:overview}

As described in the preceding section, ASV models were observed to exhibit vulnerabilities when confronted with complex scenarios. However, in real-world settings, ASV systems often encounter these threats simultaneously. Therefore, it is imperative to develop a robust system that can effectively counter these challenges in a concurrent manner.

\begin{figure}[tbp]
\centering
\includegraphics[width=8cm]{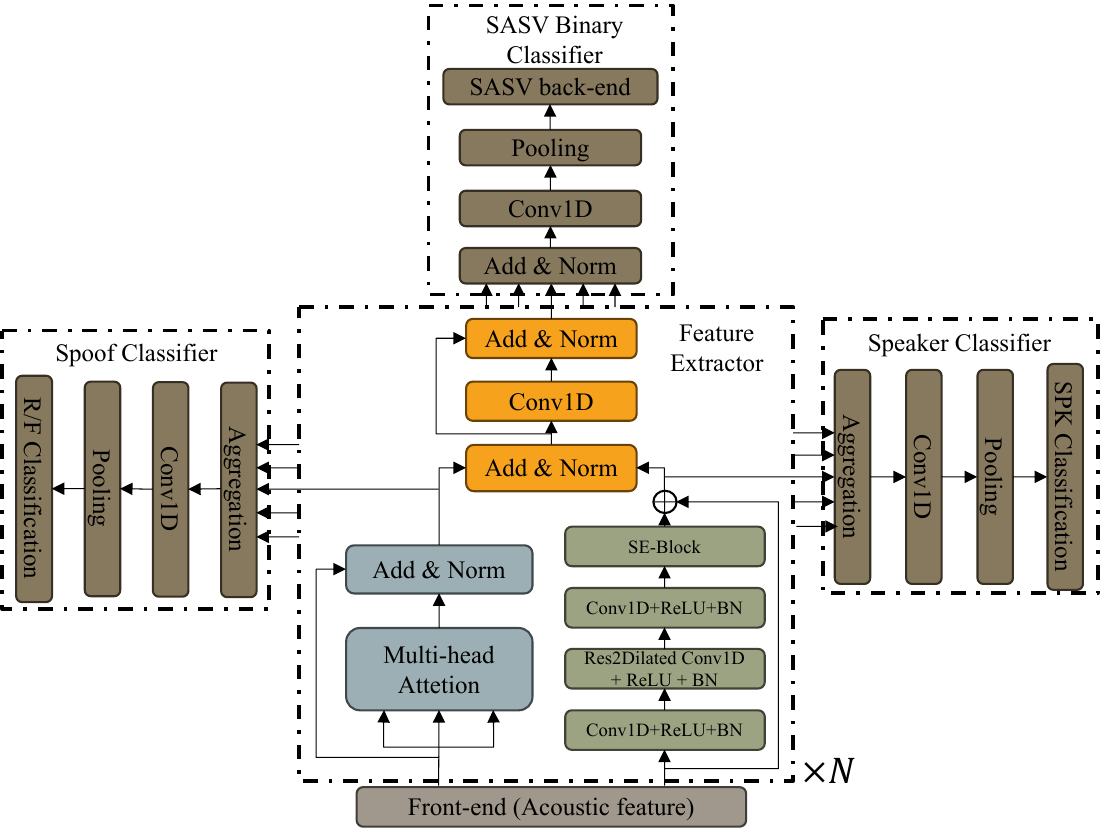}
\caption{Architecture of the dual-path model consisting of four submodules: a generic feature extractor, a speaker classifier, a spoof classifier, and a SASV binary classifier. Notably, the feature extractor adopts an asymmetric dual-path structure.}
\label{fig:arch-dual-path}
\end{figure}

In this section, building upon the foundation laid by the previous studies \cite{attention-backend,sasv-attention-backend,robustmaml}, we present an integrated solution designed to address these challenges comprehensively. The proposed solution utilizes an asymmetric dual-path model equipped with a multi-task learning strategy, as depicted in Figure \ref{fig:arch-dual-path}. This model is tasked with simultaneously handling the ASV task, the anti-spoofing task, and the spoofing-aware ASV task. Furthermore, we incorporate pair-wise learning and spoofing attack simulation into the meta-learning paradigms to train the model so that the model is equipped with the capability to address channel mismatch, spoofing attacks, and domain mismatch issues.

In comparison to traditional robust systems targeting a single threat, the proposed approach presents two distinct advantages. Firstly, it capitalizes on the interactions between the spoofing-aware attention back-end and the meta-learning-based generalization, resulting in the enhancement of both subsystems. Secondly, the end-to-end training methodology enables more efficient signal propagation to train a system robust against multiple risks. The initial gains observed in individual tasks in the previous studies \cite{attention-backend,sasv-attention-backend,robustmaml} clearly underscore the promising potential of this integrated approach.

\subsection{Model architecture and objectives}
\label{sec:model}

Figure \ref{fig:arch-dual-path} provides a visual representation of the architecture of the proposed spoofing-aware ASV model. This model consists of four main components:
\begin{itemize}
    \item Asymmetric dual-path feature extractor
    \item Speaker classifier
    \item Spoof classifier
    \item SASV binary classifier
\end{itemize}
Each of these components will be elaborated upon in detail in the subsequent section.

\subsubsection{Asymmetric dual-path feature extractor}

As depicted in Figure \ref{fig:arch-dual-path}, the asymmetric dual-path feature extractor comprises N blocks, each with distinct parameters but an identical architecture. Within each block, two different branches share the same input, denoted as $\boldsymbol{X}$. The left multi-head attention branch, inspired by \cite{rawformer,safeear,li2024audio,li2024iianet}, primarily focuses on extracting features for the anti-spoofing task. Conversely, the right SE-Res2Block, proposed in \cite{ecapa-tdnn}, is primarily responsible for feature extraction for the ASV task. The outputs of these two branches are combined through addition and layer normalization operations. The computational process can be expressed as follows:
\begin{align}
    \boldsymbol{H}_{left}^{n} & = f_{left}(\boldsymbol{X}), \\
    \boldsymbol{H}_{right}^{n} & = f_{right}(\boldsymbol{X}), \\
    \boldsymbol{H}_{fuse}^{n} & = \text{LN}(\boldsymbol{H}_{spoof}^{n} + \boldsymbol{H}_{asv}^{n}),
\end{align}
where $\boldsymbol{H}_{left}^{n}$, $\boldsymbol{H}_{right}^{n}$, and $\boldsymbol{H}_{fuse}^{n}$ represent the outputs of the left branch, right branch, and the fusion operation, respectively, within the $n$-th block. The functions $f_{left}(\cdot)$ and $f_{right}(\cdot)$ correspond to the transformations in the two branches, while $\text{LN}(\cdot)$ signifies the layer normalization, identical to the one used in \cite{attention-all-you-need}.

Furthermore, a one-dimensional convolutional layer was employed to operate on the output of the fusion operation, enhancing the non-linearity of the entire block. In contrast to the use of fully connected layers in \cite{attention-all-you-need}, we contend that the local correlations in speech features need to be emphasized. Thus, drawing inspiration from the operations described in FastSpeech \cite{fastspeech,fastspeech2}, convolutional layers were incorporated to capture these local correlations. Additionally, residual connections and layer normalization were utilized to facilitate more stable training of the model. The entire computational process is illustrated as follows:
\begin{align}
    \boldsymbol{H}_{final}^{n}=\text{LN}((\text{ReLU}(\text{Conv}(\boldsymbol{H}_{fuse}^{n})) + \boldsymbol{H}_{fuse}^{n}),
\end{align}
where $\boldsymbol{H}_{final}^{n}$ represents the final output of the $n$-th block.

\subsubsection{Speaker classifier and spoof classifier}

The architectures of both the speaker classifier and the spoof classifier in the model are nearly identical, differing only in the output nodes of the final fully connected layers, which are associated with the number of classes. Drawing inspiration from the successful utilization of multi-layer feature aggregation as seen in \cite{ecapa-tdnn} and \cite{mfa-conformer}, we have also incorporated this strategy to aggregate the output of each left branch and each right branch for the anti-spoofing and ASV sub-objectives, respectively. This aggregation can be expressed as follows:
\begin{align}
    \boldsymbol{H}_{spoof} & = \frac{1}{N}\sum_{n=1}^N\boldsymbol{H}_{left}^n, \\
    \boldsymbol{H}_{asv} & = \frac{1}{N}\sum_{n=1}^N\boldsymbol{H}_{right}^n,
\end{align}
Here, $\boldsymbol{H}_{asv}$ and $\boldsymbol{H}_{spoof}$ represent the aggregated output for the ASV and anti-spoofing tasks, respectively.

Both of the aggregated features undergo transformation through a one-dimensional convolutional layer followed by an average pooling layer. Subsequently, the classification probabilities are computed based on the outputs $\boldsymbol{e}_{asv}$ and $\boldsymbol{e}_{spoof}$ from the pooling layers. This computation can be expressed as follows:
\begin{align}
    P(\boldsymbol{e}_{spoof}) & = \frac{1}{1+\exp(-\boldsymbol{e}_{spoof}^\top\boldsymbol{\theta}_{spoof})}, \\
    \mathcal{L}_{spoof} & = -\frac{1}{M}\sum_{m}^{M}[\mathcal{I}(y^{m}_{spoof} = 1)\log P(\boldsymbol{e}^{m}_{spoof}) \nonumber \\
    & + \mathcal{I}(y^{m}_{spoof} \neq 1)\log (1 - P(\boldsymbol{e}^{m}_{spoof}))], \\
    P(\boldsymbol{e}_{asv}, s_i) & = \frac{\exp(h_{\boldsymbol{\theta}_{asv}^{s_i}}(\boldsymbol{e}_{asv}))}{\sum_{j=1}^{S}\exp(h_{\boldsymbol{\theta}_{asv}^{s_j}}(\boldsymbol{e}_{asv}))}, \label{ch6:eq:softmax} \\
    \mathcal{L}_{asv} & = -\frac{1}{M}\sum_{m}^{M} \log P(\boldsymbol{e}_{asv}^m, s_{i}^m),
\end{align}
where $P(\boldsymbol{e}_{spoof})$ and $P(\boldsymbol{e}_{asv})$ represent the probabilities for the anti-spoofing and ASV tasks, respectively. $\mathcal{I}(\cdot)$ is an indicator function that returns 1 if the condition is true and 0 otherwise. The variable $S$ denotes the number of speakers in the training dataset, while $M$ represents the number of training samples. $s_i^m$ signifies the $m$-th training sample with the $i$-th speaker in the training dataset. Furthermore, $\boldsymbol{\theta}_{spoof}$ and $\boldsymbol{\theta}_{asv}$ correspond to the parameters of the output layers, respectively.

\subsubsection{SASV binary classifier}

For the SASV binary classifier, the initial step involves aggregating the outputs $\boldsymbol{H}_{fuse}$ from each block by summation. Subsequently, layer normalization is applied to the aggregation, as expressed in the following equation:
\begin{align}
    \boldsymbol{H}_{fuse} & = \text{LN}(\sum_{n=1}^N\boldsymbol{H}_{final}^n),  
\end{align}
Here, $\boldsymbol{H}_{fuse}$ represents the output of the aggregation. Following this, the aggregated features undergo transformation through a one-dimensional convolutional layer and a pooling layer, generating an embedding denoted as $\boldsymbol{e}_{fuse}$.

The architecture employed for the sub-module labeled ``SASV back-end'' in Figure \ref{fig:arch-dual-path} is identical to that outlined in \cite{sasv-attention-backend}. However, our approach is distinct due to the assertion that the informative embedding $\boldsymbol{e}_{fuse}$ can serve both the anti-spoofing and ASV tasks. Consequently, this embedding is utilized for both CM decision and ASV decision, which diverges from the input used in the SASV back-end, as detailed in \cite{sasv-attention-backend}.

\subsection{Learning paradigm}
\label{sec:paradigm}

To confer resilience upon the proposed model, we integrate the pair-wise learning paradigm, simulation of spoofing attacks, and meta-learning paradigm. As elucidated in \cite{zc-meta-grl}, meta-learning can be understood as a process of bilevel optimization involving a hierarchical optimization problem where one optimization contains another as a constraint \cite{bilevel1,bilevel2}. This bilevel optimization is divided into two loops, the outer and inner, which effectively simulate domain mismatch during the training of machine learning models. It can be formalized as follows:
\begin{align}
    \theta_{i + 1} & = \operatorname*{argmin}_{\theta} \sum_{i}^{T} \mathcal{L}^{outer}(\theta_{i}^{*}, \mathcal{D}_{source}^{meta-test(i)}) \nonumber \\
     & \;\;\;\; + \mathcal{L}^{inner}(\theta_{i}, \mathcal{D}_{source}^{meta-train(i)}) \\
    \text{s}.\text{t}. \quad \theta_{i}^{*} & = \operatorname*{argmin}_{\theta} \mathcal{L}^{inner}(\theta_{i}, \mathcal{D}_{source}^{meta-train(i)}),
    \label{eq:station}
\end{align}
where $\mathcal{D}$ represents the training dataset from the source domain, which is divided into a meta-train dataset denoted as $\mathcal{D}_{source}^{meta-train}$ and a meta-test dataset denoted as $\mathcal{D}_{source}^{meta-test}$. The combination of a meta-train and a meta-test dataset constitutes a meta-task, indexed by $i$, which means the $i$-th one in total $T$ meta-tasks. $\theta_i$ and $\theta_{i+1}$ correspond to the parameters of the inner and outer loops, respectively. Note that there exists a leader-follower asymmetry between the outer and inner loops.

\setlength{\tabcolsep}{3.0mm}
\begin{table}[tbp]
\setlength\tabcolsep{3.5pt}
  \caption{Differences in EER between the proposed approach and the baseline. The first row of the table denotes genres of testing utterances, and the first column of the table represents genres of enrollment utterances. The \textcolor{blue}{blue} number means the proposed approach outperforms the baseline. The \textcolor{red}{red} number means the baseline outperforms the proposed approach.}
  \vspace{2mm}
  \label{tab:result_1}
  \centering
  \begin{tabular}{r c c c c c c c c}
    \toprule
   & \multicolumn{1}{c}{\textbf{dr}} & \multicolumn{1}{c}{\textbf{en}} & \multicolumn{1}{c}{\textbf{in}} & \multicolumn{1}{c}{\textbf{lb}} & \multicolumn{1}{c}{\textbf{re}} & \multicolumn{1}{c}{\textbf{si}} & \multicolumn{1}{c}{\textbf{sp}} & \multicolumn{1}{c}{\textbf{vl}} \\
    \midrule
    
    dr & +1.65 & \textcolor{red}{-1.23} & \textcolor{blue}{+1.72} & \textcolor{blue}{+5.97} & -      & -     & -     & - \\
    en & \textcolor{blue}{+2.71} & +3.06 & \textcolor{red}{-0.02} & \textcolor{red}{-1.04} & \textcolor{red}{-7.35} & \textcolor{blue}{+3.83} & \textcolor{blue}{+1.20} & \textcolor{red}{-1.06} \\
    in & \textcolor{blue}{+4.32} & \textcolor{blue}{+2.13} & +2.49 & \textcolor{red}{-0.64} & \textcolor{blue}{+2.70} & \textcolor{blue}{+2.72} & \textcolor{blue}{+3.94} & \textcolor{blue}{+14.06} \\
    lb & \textcolor{red}{-1.74} & \textcolor{red}{-1.13} & \textcolor{red}{-2.91} & +0.02 & - & \textcolor{blue}{+7.73} & - & \textcolor{blue}{+6.71}                \\
    sp & \textcolor{blue}{+2.98} & \textcolor{blue}{+0.36} & \textcolor{blue}{+3.34} & \textcolor{blue}{+5.17} & - & \textcolor{blue}{+2.07} & -0.79 & - \\
    vl & \textcolor{red}{-6.52} & \textcolor{blue}{+0.51} & \textcolor{blue}{+4.75} & \textcolor{blue}{+2.63} & - & \textcolor{red}{-2.92} & - & +2.01           \\
    \bottomrule
  \end{tabular}
\end{table}

\begin{table}[tbp]
\footnotesize
\setlength\tabcolsep{0.8mm}
  \caption{Results of the proposed system on CNCeleb.Eval and CNComplex testing datasets. The performance is evaluated under the metrics of SV-EER and SASV-EER metrics, respectively.}
  \vspace{2mm}
  \label{tab:result3}
  \centering
  \begin{tabular}{lccccc}
    \toprule
    \multirow{2}{*}{\textbf{Protocol}} & \multirow{2}{*}{\textbf{Training dataset}} & \multicolumn{2}{c}{\textbf{CNCeleb.Eval}} & \multicolumn{2}{c}{\textbf{CNComplex}} \\
    \cmidrule(l{0em}r{0em}){3-6}
    & & SV-EER & SASV-EER & SV-EER & SASV-EER \\
    \midrule
    \multirow{2}{*}{\textbf{CGP I}}   & CNCeleb 1\&2             & 7.96  & - & 8.52 & 7.37  \\
                                      & Combination   & 7.79  & - & 7.56 & 7.25  \\
    \toprule
    \multirow{2}{*}{\textbf{CGP II}}  & CNCeleb 1\&2             & 8.24 & - & 8.85 & 8.57  \\
                                      & Combination   & 7.96 & - & 8.34 & 8.47  \\
    \toprule
    \multirow{2}{*}{\textbf{CGP III}} & CNCeleb 1\&2             & 8.43 & - & 9.07 & 8.52  \\
                                      & Combination   & 8.19 & - & 8.82 & 8.25  \\
    \toprule
    \multirow{2}{*}{\textbf{CGP IV}}  & CNCeleb 1\&2             & 8.23 & - & 8.68 & 7.73  \\
                                      & Combination   & 8.13 & - & 8.45 & 7.48  \\
    \bottomrule
  \end{tabular}
\end{table}

\begin{table*}[tbp]
  \caption{EER (\%) of experimental results on CNCeleb.Eval testing dataset for the scenario of domain mismatch. For each protocol, the baseline system is established using the proposed model trained through the straightforward supervised learning paradigm. A bold number indicates the highest performance of this genre. The presence of an unseen group is indicated within brackets}.
  \label{tab:exp_result3}
  \vspace{2mm}
  \setlength\tabcolsep{2mm}
  \centering
  \begin{tabular}{ccccccccccccc} 
    \toprule
    \multirow{2}{*}{\textbf{Protocol}} & \multirow{2}{*}{\textbf{System}} & \multirow{2}{*}{\textbf{Overall}} & \multicolumn{3}{c}{\textbf{Group I}} & \multicolumn{3}{c}{\textbf{Group II}} & \multicolumn{2}{c}{\textbf{Group III}} & \multicolumn{2}{c}{\textbf{Group IV}} \\
    \cmidrule(l{0em}r{0em}){4-13}
    &  &  & \textbf{dr} & \textbf{vl} & \textbf{sp} & \textbf{en} & \textbf{in} & \textbf{pl} & \textbf{lb} & \textbf{mo} & \textbf{si} & \textbf{re} \\
    \toprule
    \multirow{1}{*}{\textbf{CGP I}} & Baseline & 21.31 & 23.15 & 20.77 & 15.86 & 21.61 & 22.99 & 27.35 & 17.50 & 29.32 & 28.63 & 14.58 \\
    \multirow{1}{*}{(Group IV)} & proposed approach & 17.42 & \textbf{22.66} & \textbf{16.44} & \textbf{14.46} & \textbf{20.95} & \textbf{19.47} & \textbf{20.93} & \textbf{16.31} & \textbf{24.57} & \textbf{22.33} & \textbf{13.89} \\
    \toprule 
    \multirow{1}{*}{\textbf{CGP II}} & Baseline  & 22.10 & 22.66 & 20.64 & 16.41 & 23.56 & 24.08 & 27.29 & 19.43 & 31.55 & 25.25 & \textbf{13.17} \\
    \multirow{1}{*}{(Group III)} & proposed approach & 18.48 & \textbf{18.04} & \textbf{18.08} &  \textbf{13.84} & \textbf{19.11} & \textbf{20.10} & \textbf{19.71} & \textbf{17.84} & \textbf{27.16}  & \textbf{24.16} & 13.91 \\
    \toprule
    \multirow{1}{*}{\textbf{CGP III}} & Baseline & 23.46 & 25.42 & 23.89 & 17.58 & 26.38 & 25.34 & 29.25 & 21.72 & 31.14 & 27.74 & \textbf{13.10} \\
    \multirow{1}{*}{(Group II)} & proposed approach & 20.02 & \textbf{23.27} & \textbf{20.70} & \textbf{16.79} & \textbf{22.08} & \textbf{22.83} & \textbf{22.96} & \textbf{19.54} & \textbf{28.53} & \textbf{22.92} & 13.55 \\
    \toprule
    \multirow{1}{*}{\textbf{CGP IV}} & Baseline & 22.59 & 24.13 & 23.47 & 16.29 & \textbf{19.50} & 22.95 & 27.81 & 19.99 & 28.12 & 24.87 & 11.33 \\
    \multirow{1}{*}{(Group I)} & proposed approach & 19.98 & \textbf{22.35} & \textbf{19.44} & \textbf{14.58} & 21.34 & \textbf{19.62} & \textbf{20.87} & \textbf{17.83} & \textbf{24.24} & \textbf{23.53} & \textbf{10.55} \\
    \bottomrule
  \end{tabular}
\end{table*}

\subsubsection{Outer loop}

In the outer loop, the genre sampling strategy is employed to simulate the domain mismatch between the meta-train and meta-test data. Specifically, we configure the number of genres in the meta-train dataset ($G_{mtr}$) as $G - 2$, with the remaining genre reserved for the meta-test dataset in all experiments in this paper. This approach is consistently applied across different training protocols.

\subsubsection{Inner loop}

Within the inner loop, we employ the training trial sampling method, as outlined in \cite{attention-backend,sasv-attention-backend}, to simulate both spoofing attacks and channel mismatch scenarios concurrently. Given that the training dataset contains both genuine and copy-synthesized utterances, with each utterance being associated with a specific genre indicating the acoustic environment, it becomes possible to leverage the pair-wise learning paradigm for generating a substantial number of training trials from the training dataset. These training trials effectively simulate various distinct scenarios involving channel mismatch and spoofing attacks, allowing the proposed model to acquire robustness-related knowledge through the inner loop training process.

\section{Experimental results}
\label{sec:exp}

\subsection{Dataset}

In the experiments, we adhere to the cross-genre training protocols outlined in \cite{zc-meta-grl} to segregate the fusion of the CNSpoof and CNCeleb 1\&2 datasets. Under each protocol, we exclude two or three genres, treating them as unseen genres found in the testing dataset. The resulting combination encompasses over 1.2 million utterances spanning eleven distinct genres. Furthermore, the diversity of the training data is enhanced by incorporating the MUSAN \cite{musan} and RIRs \cite{rirs} datasets.

CNCeleb.Eval and CNComplex are both employed as the testing datasets, along with their respective evaluation protocols, to assess the performance of the proposed model. Both datasets feature 196 speakers for enrollment and consist of 17,777 utterances in the testing data, resulting in a total of 3,484,292 testing trials for each evaluation protocol.

\subsection{Experimental settings}

To facilitate the implementation of the meta-learning paradigm, a subset of 64 samples was randomly selected from the dataset. Within this subset, one genre was chosen randomly to constitute the meta-test dataset for the outer loop, while the remaining samples were utilized as the meta-train dataset for the inner loop. All systems underwent training for 80 epochs using an Adam optimizer \cite{adam}. Training commenced with an initial learning rate of 0.01, $\beta_{1}$ of 0.9, and $\beta_2$ of 0.98. A warm-up strategy was employed, gradually increasing the learning rate from 0 to 0.001 within the first 5,000 steps and subsequently reducing the learning rate using an exponential scheduler.

\subsection{Results and analysis}

To illustrate the promising outcomes of the proposed model, which integrates a comprehensive learning paradigm, we conduct independent evaluations of the system across various scenarios encompassing channel mismatch, spoofing attacks, and domain mismatch, respectively.

\subsubsection{Result for channel mismatch scenario}

To assess the performance of the proposed approach in scenarios involving channel mismatch, we utilize a protocol based on the CNCeleb.Eval testing dataset similar to the one outlined in \cite{attention-backend}, which encompasses various mismatch conditions. In this experimental evaluation, the baseline system is established using the proposed model trained through the straightforward supervised learning paradigm without the meta-learning paradigm. The differences in EER between the system and the baseline system within the context of channel mismatch are presented in Table \ref{tab:result_1}. Each cell in the table indicates the EER difference. 

The table reveals that the proposed approach exhibits significantly improved performance in scenarios characterized by channel mismatch, which refers to the genre mismatch between the enrollment and testing utterances in the case. Positive differences are notable when testing utterances from the ``dr,'' ``in,'' ``lb,'' ``sp,'' and ``vl'' genres are enrolled from the ``dr,'' ``en,'' ``in,'' ``sp,'' and ``vl'' genres, respectively. This analysis demonstrates that the proposed approach is particularly effective in mitigating the impact of channel mismatch, emphasizing its capability to adapt to varied mismatch conditions during channel variability scenarios.

\subsubsection{Result for spoofing attack scenario}

Table \ref{tab:result3} provides a comprehensive overview of the performance evaluation results for the proposed system considering various training protocols and two separate testing datasets. The evaluation is conducted using the SV-EER and SASV-EER metrics.

First and foremost, comparing the number of SV-EERs on the CNCeleb.Eval dataset of two different training datasets for all training protocols, it is evident that the inclusion of the CNSpoof dataset in the training data consistently improves the performance of the traditional ASV task, aligning with the results presented in Section \ref{sec:preliminary_result}. In all training protocols, models trained with the CNSpoof data consistently outperform those trained without it, as evidenced by improvements in all SV-EER numbers.

Furthermore, it is noteworthy that the SASV-EER values for the same system demonstrate comparable results across different protocols when compared to the SV-EER. This observation underscores the system's capacity for maintaining stability in addressing spoofing attacks, which is a crucial attribute for practical deployment scenarios.

\subsubsection{Result for domain mismatch scenario}

Table \ref{tab:exp_result3} shows the EER results for the scenario of domain mismatch under various protocols. Notably, regardless of the training protocols employed, the proposed model, trained using the integrated learning paradigm, exhibits a substantial overall improvement in performance, as indicated by the third column in the table.

Within each protocol, the presence of an unseen group is indicated within brackets. The proposed approach greatly outperformed in terms of EER in comparison to the baseline system, particularly in the ``si'' genre of CGP I and the ``pl'' genre of CGP III. This observation underscores the effectiveness of the proposed model, trained using the integrated learning paradigm, in mitigating the challenges posed by domain mismatch in the context of the ASV task.

Furthermore, the EER numbers for seen genres are also enhanced through the use of the integrated learning paradigm. This suggests that the gradients derived from the meta-test optimization process can effectively serve as a regularization term, correcting the biases inherently introduced by the supervised learning paradigm.

In summary, the results presented in the table highlight that the proposed approach consistently outperforms the baseline system in mitigating domain mismatch, resulting in improved overall EER and EER reductions across various genres in each training protocol.

\section{Conclusion}
\label{sec:conclusion}

In this paper, we proposed an asymmetric dual-path model and an integrated framework unifying pair-wise learning, spoofing simulation, and meta-learning for robustness against channel mismatch, spoofing attacks, and domain shifts. The bilevel optimization structure makes it possible to these multifaceted issues concurrently in an end-to-end model.

To demonstrate the advantages of the proposed model and learning paradigm, a series of experiments were conducted to show the performance improvement independently under scenarios of channel mismatch, spoofing attacks, and domain mismatch, respectively. The experiments demonstrated consistent improvements over baseline systems trained in a supervised learning paradigm. The proposed approach provides a blueprint for developing production ASV systems that are dependable under diverse channel mismatch, spoofing attacks, and domain mismatch.

\section{ACKNOWLEDGMENTS}
\label{sec:ack}
This study is partially supported by JST CREST Grants (JPMJCR18A6, JPMJCR20D3) and JST AIP Acceleration Research (JPMJCR24U3). This research was also partially funded by the project to develop and demonstrate countermeasures against disinformation and misinformation on the Internet with the Ministry of Internal Affairs and Communications of Japan.


\bibliographystyle{IEEEbib}
\bibliography{strings,refs}

\end{document}